\documentclass[useAMS,usenatbib]{mn2e}
\usepackage{graphicx}
\usepackage{amssymb}

\title[A universal scaling for short and long GRBs]{A universal scaling for short and long gamma-ray bursts: $E_{X,iso}-E_{\gamma,iso}-E_{pk}$}

\author[M.G. Bernardini et al.]{M.G. Bernardini$^{1}$\thanks{E-mail: grazia.bernardini@brera.inaf.it}, R. Margutti$^{2,1}$, E. Zaninoni$^{1,3}$, G. Chincarini$^{1,4}$\\
$^{1}$INAF - Osservatorio Astronomico di Brera, via Bianchi 46, I-23807 Merate (LC), Italy \\
$^{2}$Harvard-Smithsonian Center for Astrophysics, 60 Garden Street, Cambridge, MA02138, USA\\
$^{3}$University of Padova, Physics \& Astronomy Dept. Galileo Galilei, via Marzolo 8, I-35131 Padova, Italy\\
$^{4}$University of Milano Bicocca, Physics Dept., p.zza della Scienza 3, I-20126 Milano, Italy}
\date{\today}

\pagerange{\pageref{firstpage}--\pageref{lastpage}} \pubyear{2012}

\begin{document}

\label{firstpage}

\maketitle

\begin{abstract}
The comprehensive statistical analysis of \emph{Swift} X-ray light-curves, collecting data from six years of operation, revealed the existence of a universal scaling among the isotropic energy emitted in the rest frame $10-10^4$ keV energy band during the prompt emission ($E_{\gamma,iso}$), the peak of the prompt emission energy spectrum ($E_{pk}$), and the X-ray energy emitted in the $0.3-10$ keV observed energy band ($E_{X,iso}$). In this paper we show that this three-parameter correlation is robust and does not depend on our definition of $E_{X,iso}$. It is shared by long, short, and low-energetic GRBs, differently from the well-known $E_{\gamma,iso}-E_{pk}$ correlation. We speculate that the ultimate physical property that regulates the GRB properties is the outflow Lorentz factor.
\end{abstract}


\section{Introduction}

The BATSE observations of the Gamma-ray Burst (GRB) prompt emission and the direct measurement of the distance provided by \emph{Beppo}SAX allowed the discovery of a correlation between the isotropic energy emitted during the prompt emission ($E_{\gamma,iso}$) and the peak of the prompt emission energy spectrum ($E_{pk}$) by \citet{2002A&A...390...81A}. Whichever the origin, the correlation itself proved to be a useful cosmological tool \citep{2008MNRAS.391..577A}.

With the advent of the \emph{Swift} satellite \citep{2004ApJ...611.1005G} it was possible to accumulate a large amount of information about the properties of GRBs. Moreover, \emph{Swift} allowed the prompt localisation of a large number of GRBs with arcsec precision, an essential information to re-point optical telescopes and determine the redshift of the sources. The $E_{\gamma,iso}-E_{pk}$ correlation has been confirmed and extended to a larger sample in the \emph{Swift} era  \citep{2006MNRAS.372..233A,2009A&A...508..173A}. Several correlations between prompt emission quantities \citep{2004ApJ...609..935Y,2004ApJ...616..331G,2006MNRAS.370..185F,2006MNRAS.369L..37L}, afterglow quantities \citep{2008MNRAS.391L..79D,2010ApJ...722L.215D}, and between prompt and afterglow emission \citep{2011MNRAS.tmp.1929D,2011arXiv1103.3978X} have been derived. In particular, the discovery of the extragalactic origin of short GRBs by \emph{Swift} \citep{2005Natur.437..851G} showed that short GRBs are outliers of the $E_{\gamma,iso}-E_{pk}$ correlation.

During the comprehensive statistical analysis of \emph{Swift} X-ray light-curves, presented by Margutti et al. (2012) as a companion paper (hereafter M12), collecting data from six years of operation of the \emph{Swift}/X-ray Telescope \citep[XRT;][]{2005SSRv..120..165B}, we proved that a universal scaling exists among $E_{\gamma,iso}$, $E_{pk}$, and the X-ray energy emitted in the $0.3-10$ keV observed energy band $E_{X,iso}$. In this paper we show that this three-parameter correlation is robust and is not dependent on our definition of $E_{X,iso}$. Differently from the well-known $E_{\gamma,iso}-E_{pk}$ correlation, long \emph{and} short GRBs share the same relation, suggesting that the physical origin of the correlation lies in what is common to the two classes, and likely independent of the progenitors and environment since both are thought to be different.

In Sect.~\ref{sect_corr} we detail the analysis of the correlation and illustrate the relevant cases and possible biases. In Sect.~\ref{discussion} we discuss our findings. Then we draw our main conclusions. We adopt standard values of the cosmological parameters: $H_\circ=70$ km s$^{-1}$ Mpc$^{-1}$, $\Omega_M=0.27$ and $\Omega_{\Lambda}=0.73$.  Errors are given at $1-\sigma$ confidence level unless otherwise stated.

\section{The $E_{X,iso}-E_{\gamma,iso}-E_{pk}$ correlation}\label{sect_corr}

\begin{figure*}
\centering
\includegraphics[width=\hsize,clip]{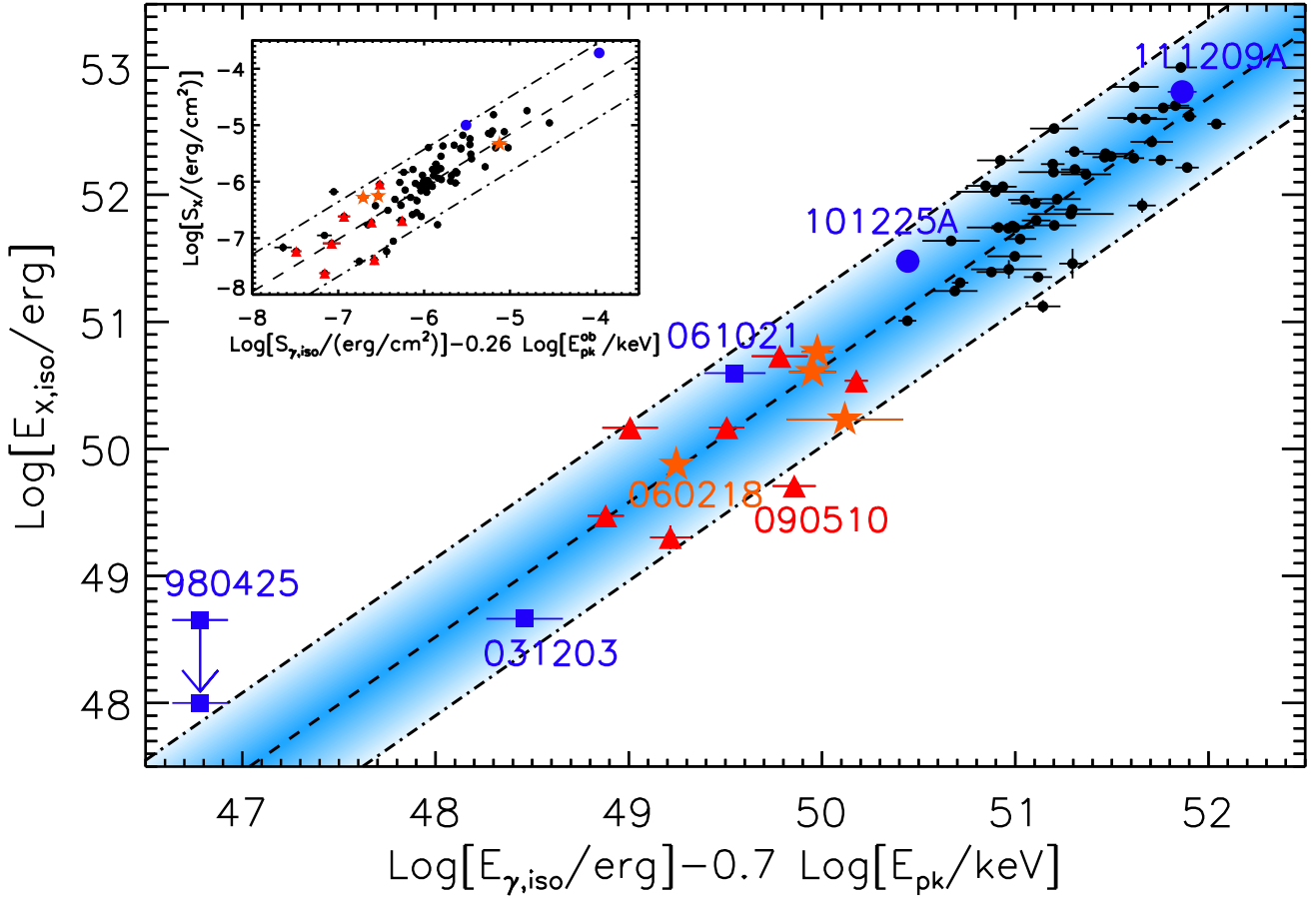}
\caption{$E_{X,iso}-E_{\gamma,iso}-E_{pk}$ correlation for the sample of $52$ LGRBs (black points) and $7$ SGRBs (red triangles). The orange stars correspond to low-energetic GRBs (GRB050416A, GRB060218, GRB060614, GRB081007). The blue squares correspond to GRB090425, GRB031203, and GRB061021, outliers of the $E_{\gamma,iso}$-$E_{pk}$ correlation. The blue circles correspond to GRB101225A and GRB111209A. The black dashed line is the best-fitting function $y=1.06(x-0.7z)-2.36$ and the blue area marks the $2-\sigma$ region. \textbf{Inset}: $S_{X}-S_{\gamma,iso}-E_{pk}^{ob}$ correlation for the sample of $71$ LGRBs (black points) and $8$ SGRBs (red triangles). The orange stars correspond to low-energetic GRBs and the blue circles correspond to GRB101225A and GRB111209A. The black dashed line is the best-fitting function $y=0.93(x-0.26z)-0.51$ and the dot-dashed lines mark the $2-\sigma$ region.}
\label{Correlation}
\end{figure*}

We considered all the GRBs for which it was possible to measure the rest frame peak energy ($E_{pk}$) and prompt emission isotropic energy in the rest frame $1-10^4$ keV energy band ($E_{\gamma,iso}$; \citealt{2006MNRAS.372..233A,2008MNRAS.391..577A,2009A&A...508..173A} and Amati, private communication). Among these, we selected those GRBs observed by \emph{Swift}/XRT that have a \emph{complete} light curve (LC), i.e. promptly re-pointed by \emph{Swift}/XRT and for which the observation was not limited by any observing constraint (see M12 for details)\footnote{This interval corresponds to $300<t\lesssim4\times 10^5$ s}. The GRBs satisfying the above criteria are $59$ (see Table~\ref{tab_sample}); among these seven are short GRBs (SGRB; hereafter, LGRB stands for long GRB).

\begin{table}
\tiny
 \begin{minipage}{85mm}
  \caption{List of the $59$ GRBs included in the sample. SGRBs are marked in boldface.}
  \label{tab_sample}
    \resizebox{\textwidth}{!}{
  \begin{tabular}{l}
\hline
\hline
GRB name\\
\hline
050401, 050416A, \textbf{050724}, 050820A, 050904, 050922C, \\
051109A, \textbf{051221A}, 060115, 060124, 060206, 060218, \\
060418, 060526, 060607A, 060614, 060707, 060814, \\
060908, 060927, \textbf{061006}, 061007, 061121, \textbf{070714B}, \\
071020, \textbf{071227}, 080319B, 080319C, 080411, 080413A, \\
080413B, 080607, 080721, 080810, 080913, 080916A, \\
080928, 081007, 081008, 081028, 081118, 081222, \\
090205, 090418A, 090423, 090424, \textbf{090510}, 090516, \\
090618, 090715B, 090812, 091018, 091020, 091029, \\
091208B, 100621A, 100814A, \textbf{100816A}, 100906A \\
\hline	
\end{tabular}	}
\end{minipage}	 	 	 														
\end{table}	 

We used the best-fitting parameters of the unabsorbed luminosity LCs from M12 to derive the isotropic energy in the \emph{observer frame} $0.3-10$ keV, $E_{X,iso}$. The time integration has been performed over the observed duration of each LC: no extrapolation was applied. For those GRBs with flaring activity, the best-fitting parameters refer to the continuum underlying the X-ray LC, therefore our definition of $E_{X,iso}$ \emph{does not include flares}, if presents. For a complete description of the data analysis and the fitting procedure we refer to M12.

We searched for a three-parameter correlation between $E_{X,iso}$, $E_{\gamma,iso}$, and $E_{pk}$. We considered an intrinsic scatter $\sigma_i$ that accounts for the possible contribution of hidden variables. Markov chain Monte Carlo techniques are used in our calculations to derive the best-fitting parameters: for each Markov chain, we generate $10^5$ samples according to the likelihood function\footnote{In our analysis we used JAGS (Just Another Gibbs Sampler).  It is a program for analysis of Bayesian hierarchical models using Markov Chain Monte Carlo simulation. More informations can be found: http://mcmc-jags.sourceforge.net/}. Then we derived coefficients according to the statistical results of the samples. This yields to:
\begin{eqnarray} 
\nonumber
Log\left[\frac{E_{X,iso}}{erg}\right]=(1.06\pm0.06)Log\left[\frac{E_{\gamma,iso}}{erg}\right]-\\
-(0.74\pm0.10)Log\left[\frac{E_{pk}}{keV}\right]-(2.36\pm0.25)\, ,
\label{best-fit}
\end{eqnarray}
with $\sigma_i=(0.31\pm0.03)$. A two-dimensional rearrangement of the best-fitting relation compared to the observations is portrayed in Fig.~\ref{Correlation}. This correlation is robust, spanning four orders of magnitude in $E_{X,iso}$ and $E_{pk}$, and six orders of magnitude in $E_{\gamma,iso}$, and combines \emph{both} SGRBs and LGRBs in a common scaling. We note that the introduction of a third variable $E_{X,iso}$ reduces the scatter of the $E_{\gamma,iso}-E_{pk}$ correlation calculated for SGRBs and LGRBs (which results $\sigma_i=0.37$). The 3-parameter relation has instead a larger scatter than the Amati relation if we were to restrict our analysis to LGRBs.

The X-ray GRB LCs in Table~\ref{tab_sample} are a representative subsample of the complete LCs in M12 (see Table~1 in M12), with the exception of type 0 ($0.7\%$ in the present sample versus $12\%$ in M12). $40\%$ of the LCs in Table~\ref{tab_sample} has a plateau (type IIa and III). For those GRBs, $E_{X,iso}\sim E_{plateau}$ (see M12). Therefore, the relation between the plateau energy and the prompt emission energy found by \citet{2011MNRAS.tmp.1929D} is likely a byproduct of the more general scaling between $E_{X,iso}$ and $E_{\gamma,iso}$.

We obtained similar results also in the observer frame (fluence). We selected $79$ GRBs from M12, eight of which are SGRBs, with complete X-ray LC and prompt emission spectrum fitted with a Band function from literature (we refer to \citealt{2011A&A...530A..21N} for GRBs observed by \emph{Fermi}/GBM until March 2010, and to GCN Circulars Archive for the others). The correlation among the X-ray fluence in the $0.3-10$ keV energy band ($S_X$), the prompt emission fluence in the \emph{observer frame} $1-10^4$ keV energy band ($S_{\gamma}$), and the observed peak energy ($E_{pk}^{ob}$) is still statistically significant (see Fig.~\ref{Correlation}, inset): $Log[S_{X}]=(0.93\pm0.06)Log[S_{\gamma}]-(0.24\pm0.09)Log[E_{pk}^{ob}]-(0.51\pm0.4)$, with $\sigma_i=(0.33\pm0.03)$. This is not surprising since the $E_{X,iso}-E_{\gamma,iso}-E_{pk}$ correlation is weakly dependent on redshift. In fact $E_{X,iso}\propto E_{\gamma,iso}$, so that varying the luminosity distance produces a shift along the direction of the best fit. The only dependence on redshift comes from $E_{pk}=E_{pk}^{ob} (1+z)$.

We tested the correlation with the addition of some peculiar GRBs. We first considered GRB980425 and GRB031203, not observed by \emph{Swift}, which are known outliers of the $E_{\gamma,iso}-E_{pk}$ correlation\footnote{The consistency of GRB031203 with the $E_{\gamma,iso}-E_{pk}$ correlation is debated. In fact the $E_{pk}$ estimated from the Konus data \citep{2005NCimC..28..351U} could be much lower, based on the dust echo observed with XMMÐNewton \citep{2006ApJ...636..967W}.} (\citealt{2006MNRAS.372..233A}, see however \citealt{2006MNRAS.372.1699G}). The best possible estimate of $E_{X,iso}$ in both cases was obtained from the data available in literature\footnote{For GRB980425 we used the \emph{Beppo}SAX/WFC, XMM and \emph{Chandra} data from \citet{2004AdSpR..34.2711P}. For GRB031203 we used the XMM observations from \citet{2004ApJ...605L.101W}. We fitted in both cases the LCs with a power-law with exponential cutoff and we integrated the best-fitting LCs to obtain the emitted energy.}. We find that GRB031203 is in excellent agreement with the correlation, while GRB980425 is a factor $\sim 10$ above the extrapolation (see Fig.~\ref{Correlation}). However, the X-ray late time emission observed in GRB980425 is likely not related to the GRB but associated to Supernova SN1998bw \citep{2004ApJ...605L..97W}. For this reason we draw an upper limit: if we assume the value of $E_{\gamma,iso}$ as an upper limit on $E_{X,iso}$ for GRB980425, then it becomes consistent with the correlation within $2-\sigma$.

\citet{2012MNRAS.tmp.2475N} reported the existence of a further outlier of the $E_{\gamma,iso}-E_{pk}$ correlation: GRB061021. This GRB, observed by \emph{Swift}/XRT, was not included in our sample because its prompt emission spectrum was fitted with a power-law with an exponential cutoff, and not with a Band spectrum. However, with $E_{\gamma,iso}$ and $E_{pk}$ from \citet{2012MNRAS.tmp.2475N}, we find that it is consistent with the three-parameter correlation. Consistently, it shares the same region of the plane occupied by other low-energetic GRBs as well as by SGRBs (see Fig.~\ref{Correlation}).

GRB060218 lasted $T_{90}\sim 2100$ s in the $15-150$ keV (BAT) energy band, with an extremely soft prompt emission spectrum $E_{pk}\sim 5$ keV. The energy emitted in the $1-10$ keV band during the simultaneous observations of BAT and XRT is therefore computed both in $E_{X,iso}$ and in $E_{\gamma,iso}$. 
We then recalculated the $0.3-10$ keV energy from $t_{start}=T_{90}$: $E_{X,iso}$ is reduced by a factor $\sim 0.24$, therefore leaving GRB060218 in the $2-\sigma$ region around the best-fitting relation.

GRB090510 is the only GRB of our sample detected at very high energies $\gtrsim 1$ GeV by \emph{Fermi}/LAT. It is also a SGRB. It is consistent with the correlation at $2-\sigma$ level (see Fig.~\ref{Correlation}). On the contrary, it is an outlier of all the correlations where prompt and X-ray emission parameters are involved (see M12, Fig.~12), as well as of the $E_{\gamma,iso}-E_{pk}$ correlation \citep{2009A&A...508..173A}. From the definition of $E_{\gamma,iso}$, it does not include the high-energy emission above $10^4$ keV. If we compute it in the prompt emission energy, GRB090510 falls off the correlation. The same is true if we add it to  $E_{X,iso}$ (the high energy component has been suggested to be associated to the afterglow emission; \citealt{2010ApJ...720.1008C,2010MNRAS.409..226K,2010MNRAS.403..926G,2010ApJ...709L.146D}). Since GRB090510 is the unique event with GeV emission promptly repointed by \emph{Swift}, we have no elements to say if this is something peculiar of this GRB or this behaviour is typical of the entire class of GRBs with GeV emission.

We portrayed in Fig.~\ref{Correlation} also GRB101225A and GRB111209A. GRB101225A is an unusual GRB (its extragalactic nature has been in fact questioned \citealt{2011Natur.480...69C}), with a gamma-ray emission exceptionally long followed by a bright X-ray transient with a hot thermal component and an unusual optical counterpart. We tested its consistency with the correlation assuming the redshift estimate from a possible supernova component in the optical light curve \citep{2011Natur.480...72T} and the $E_{pk}$ from the fit of the BAT spectrum \citep{2011Natur.480...72T}. $E_{\gamma,iso}$ has been derived from the BAT energy assuming the scaling in Fig.~C3 of M12. GRB111209A is another exceptionally long event. We calculated $E_{\gamma,iso}$ and $E_{pk}$ from Konus-Wind data \citep{2011GCN..12663...1G}, and derived the $E_{X,iso}$ with the same procedure as in M12. Notably, both GRBs are consistent with the correlation.

\subsection{Caveat on the definition of $E_{X,iso}$}

In what follows we discuss the possible biasses arising from our definition of $E_{X,iso}$ (for $E_{\gamma,iso}$ and $E_{pk}$ we refer to \citealt{2006MNRAS.372..233A} and \citealt{2009A&A...508..173A}).

\begin{figure*}
\centering
\includegraphics[width=0.45 \hsize,clip]{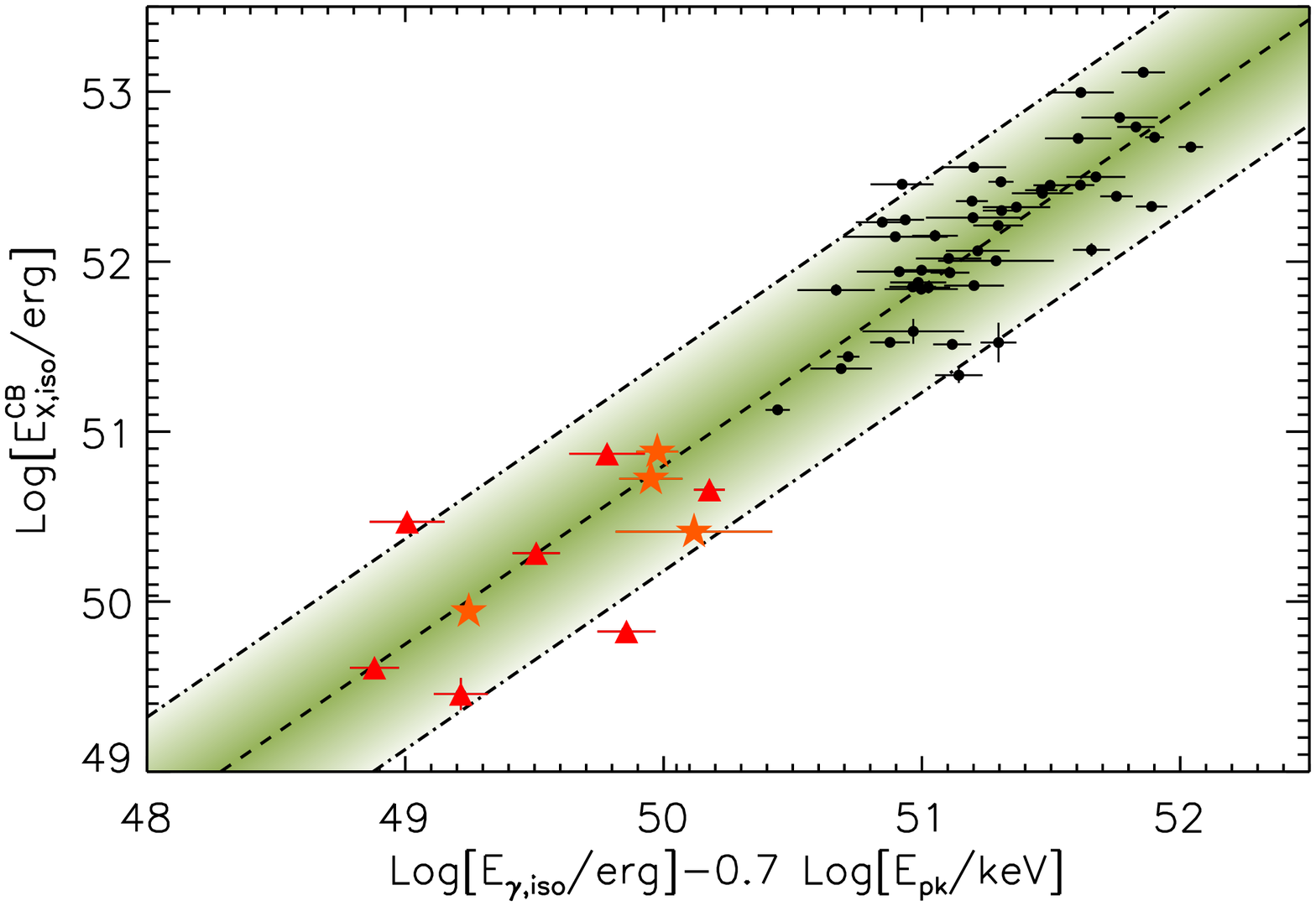}
\includegraphics[width=0.45 \hsize,clip]{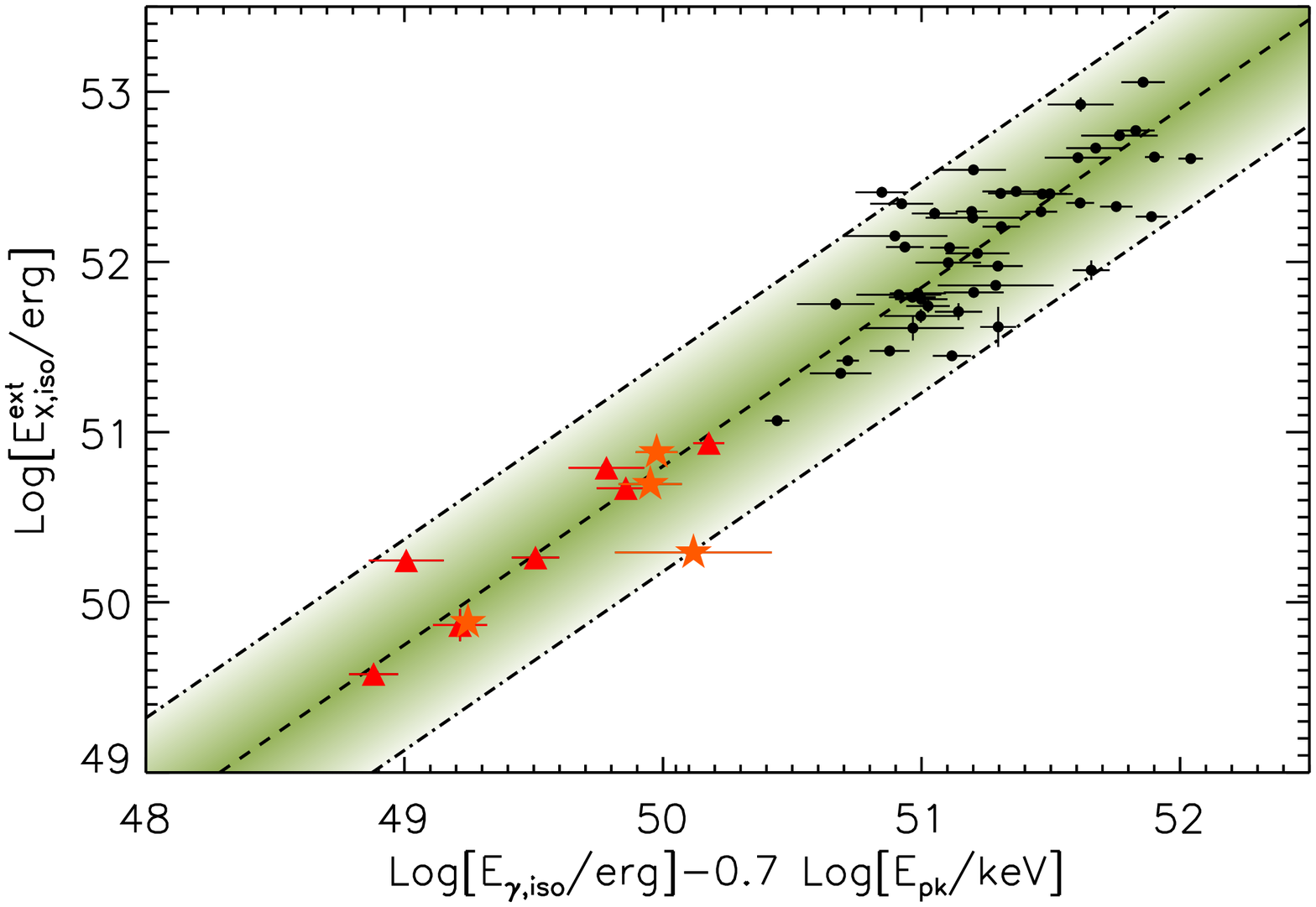}
\caption{$E_{X,iso}-E_{\gamma,iso}-E_{pk}$ correlation for the sample of $52$ LGRBs (black points) and $7$ SGRBs (red triangles): in the common cosmological rest-frame $0.3-30$ keV energy band (\emph{left panel}); for the extrapolated LCs (\emph{rigth panel}). The orange stars correspond to low-energetic GRBs. The black dashed line is the best-fitting function $y=1.05(x-0.7z)-1.7$ and the green area marks the $2-\sigma$ region.}
\label{RF_corr}
\end{figure*}

$E_{X,iso}$ has been computed in the \emph{observer frame} $0.3-10$ keV band. This means that we are sampling slightly different \emph{rest frame} energy bands for different redshifts. We therefore applied to $E_{X,iso}$ a k-correction \citep{2001AJ....121.2879B}, adopting an average spectrum $N(\nu)\propto \nu^{-\Gamma}$. To account for the spectral evolution of the X-ray LC \citep{2011arXiv1112.1058G}, we calculated the k-correction for each phase\footnote{as defined by the temporal breaks}, assuming an average spectral index $\Gamma$ from the time resolved spectral analysis presented in M12. The \emph{total} $E_{X,iso}^{CB}$ is then computed as the sum of all the contributions. The result for a common rest-frame $0.3-30$ keV energy band as an example is portrayed in Fig.~\ref{RF_corr}, \emph{left panel}. Changing the limits of the energy band does not change the result significantly. The correlation is almost unchanged: $Log[E_{X,iso}^{CB}]=(1.05\pm0.06)Log[E_{\gamma,iso}]-(0.74\pm0.10)Log[E_{pk}]-(1.70\pm0.24)$, with $\sigma_i=(0.31\pm0.03)$. We can confidently exclude that the correlation is induced by the choice of a specific energy band in the determination of $E_{X,iso}$.

Another source of arbitrariness may be the time interval over which the LCs have been integrated. The average time interval considered here is $100\lesssim t^{RF} \lesssim 10^6$ s, which is broad enough to be representative of the entire X-ray emission. However, $E_{X,iso}$ is an underestimate of the total X-ray energy. This is particularly relevant for SGRBs that are found to sample the low luminosity end of the LGRB luminosity distribution, and that decay steeper than the long ones (see M12). We calculated the correction factor by extrapolating the LCs backwards to $t_{start}^{RF}=T_{90}^{RF}$ \citep[we refer to][ for the $T_{90}$ values]{BATcatalog}, and forwards to $t_{end}^{RF}=10^7$ s. The results are shown in Fig.~\ref{RF_corr}, \emph{right panel}. As before, the total energy calculated from the extrapolated LCs $E_{X,iso}^{ext}$ strongly correlates with $E_{\gamma,iso}$ and $E_{pk}$ with the same best-fitting parameters than in Fig.~\ref{Correlation} within uncertainties. Notably, SGRB distribution is much narrower for the extrapolated light curves.

\section{Discussion}\label{discussion}

In the previous Section we showed that a universal scaling exists between prompt and X-ray parameters:
\begin{equation}
E_{X,iso}\sim\frac{E_{\gamma,iso}}{E_{pk}^{3/4}}\,.
\label{corr_ana}
\end{equation}
It is shared by SGRBs, LGRBs, and low-energetic events, with smaller scatter than the $E_{X,iso}-E_{\gamma,iso}$, $E_{X,iso}-E_{pk}$ and $E_{\gamma,iso}-E_{pk}$ correlations when including the SGRB sample. Whatever origins this correlation, this must be common to the different GRB classes, with very limited dependence on environment and progenitor properties (which are thought to be different for the different classes of bursts).

\begin{figure}
\centering
\includegraphics[width=\hsize,clip]{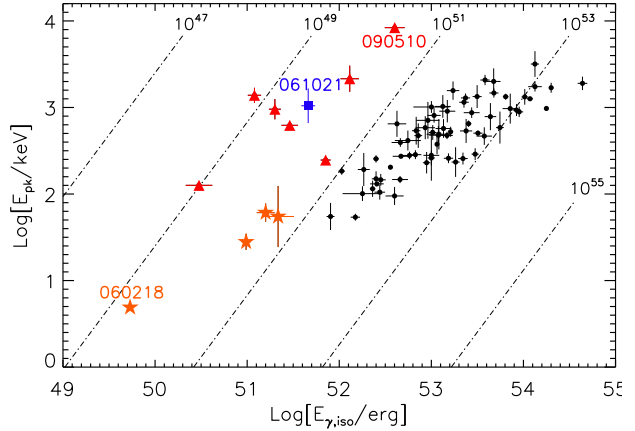}
\caption{$E_{X,iso}-E_{\gamma,iso}-E_{pk}$ correlation for the sample of $52$ LGRBs (black points) and $7$ SGRBs (red triangles) projected on the $E_{\gamma,iso}-E_{pk}$ plane. The orange stars correspond to the low-energetic GRBs and the blue square to GRB061021. The dashed-dotted lines correspond to different values of $E_{X,iso}$.}
\label{amati}
\end{figure}

At a more careful inspection we notice that SGRBs are clustered in a separate region of the plane with respect to the energetic LGRBs (see Fig.~\ref{Correlation}). This region, limited by $E_{X,iso}\sim 10^{51}$ erg, is shared with low-energetic LGRBs ($E_{\gamma,iso}\lesssim10^{51}$ erg, $E_{pk}\lesssim 50$ keV). Fig.~\ref{amati} portrays the projection of the three-parameter correlation on the $E_{\gamma,iso}-E_{pk}$ plane (see Fig.~\ref{amati}). Energetic LGRBs are characterised by the well-known $E_{pk}\propto E_{\gamma,iso}^{\alpha}$, with $\alpha\sim 0.5$ \citep{2006MNRAS.372..233A}, and their $E_{X,iso}$ spans the interval $10^{51}-10^{53}$ erg. SGRBs are not simply ``outliers'' of the Amati relation, but seem to be clustered on a different relation $E_{pk}\propto E_{\gamma,iso}^{\alpha}$, with $\alpha\gtrsim 1$, whose normalisation depends on the systematically lower $E_{X,iso}$. Interestingly, low-energetic GRBs belong to the common tail of both the LGRB and SGRB relations. 

\begin{figure}
\centering
\includegraphics[width=\hsize,clip]{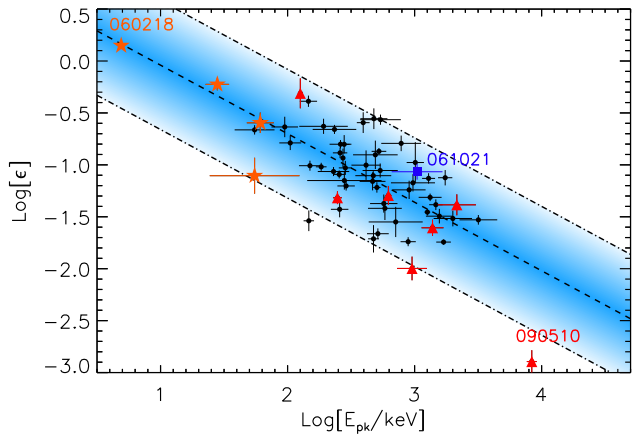}
\caption{Efficiency $\epsilon=E_{X,iso}/E_{\gamma,iso}$ vs $E_{pk}$ correlation for the sample of $52$ LGRBs (black points) and $7$ SGRBs (red triangles). The orange stars correspond to the low-energetic GRBs and the blue square to GRB061021. The black dashed line is the best-fit function $y=-(0.66\pm0.16)x+(0.62\pm0.42)$ and the blue area marks the $2-\sigma$ region, with $\sigma_i=(0.31\pm0.07)$.}
\label{eff}
\end{figure}

A convenient way to rearrange Eq.~\ref{corr_ana} is to introduce an efficiency factor between X-ray and gamma-rays, defined as the ratio between the X-ray and the prompt emission energies ($\epsilon=E_{X,iso}/E_{\gamma,iso}$). This ratio is independent from the assumption of isotropy, as far as the prompt and the X-ray emission have the same jet opening angle. The jet opening angle is expected to evolve with radius if a non-radial magnetic field plays a role in collimation and acceleration \citep[see e.g.][]{2003MNRAS.341.1360L}; if this is the case, the scatter in Fig.~\ref{eff} might be produced by different corrections for $E_{X,iso}$ and $E_{\gamma,iso}$. We conventionally define highly efficient GRBs those that have a low $\epsilon$, meaning that the majority of energy is emitted during the prompt phase. From Eq.~\ref{corr_ana} it follows that the efficiency depends, to a first approximation, only on $E_{pk}$: $\epsilon\sim E_{pk}^{-\alpha}$, with $\alpha\sim 0.7$ (see Fig.~\ref{eff}). We notice that SGRBs and low-energetic GRBs now form two separate classes: the low-energetic GRBs are less efficient, while SGRBs are as efficient as energetic LGRBs, although characterised by a lower absolute values of $E_{X,iso}$ and $E_{\gamma,iso}$.

\citet{2011arXiv1107.4096G} showed that a linear correlation exists between $E_{pk}$ and $\Gamma$, where $\Gamma$ is the Lorentz factor of the outflow \citep[see also][]{2010ApJ...725.2209L}, and concluded that different $\Gamma$ factors are responsible for the observed correlations in the prompt emission. If this is true, then the Lorentz factor is the ultimate parameter ruling the efficiency between prompt gamma-ray and late X-ray emission: $\epsilon \propto \Gamma^{-\alpha}$. This has important implications: fast ejecta will emit a large fraction of their energy budget during the prompt phase, while slow ejecta will retain large fraction of their initial energy for the following phases. Low-energetic GRBs (e.g. GRB060218) are thought to be mildly relativistic \citep{2007ApJ...667..351W,2007MNRAS.382L..77G}, and in fact they are consistently on the low efficiency tail of Fig.~\ref{eff}. GRB090510, which is two orders of magnitude more efficient than the majority of our sample, has a very high Lorentz factor ($\Gamma\gtrsim 1000$, \citealt{2011ApJ...738..138R}; $\Gamma\sim 800$, \citealt{2010MNRAS.403..926G}).

The photospheric model identifies $E_{pk}$ with the thermal peak of the photospheric emission. Its value in this interpretation is coupled to the main properties of the outflow, as the luminosity and the Lorentz factor\footnote{$E_{pk}\propto \Gamma$ is expected for magnetic dissipation and photospheric emission. On the contrary, an inverse proportionality is expected for internal shocks \citep[see][Table 1]{2002ApJ...581.1236Z}.} \citep{2005ApJ...628..847R,2006ApJ...652.1400R,2007ApJ...666.1012T}. Since similar properties are expected for the SGRB outflow \citep{2007PhR...442..166N}, it is reasonable to think that thermalisation below the photosphere occurs for those GRBs as well, and that some relation between $E_{\gamma,iso}$ and $E_{pk}$ exists also in this case \citep{2007ApJ...666.1012T}. Significant dissipation is required close to the photosphere of the flow to lead to the observed spectra. The source of such a dissipation can be internal shocks, nuclear collisions or magnetic reconnection \citep{2005ApJ...628..847R}. In the latter case, \citet{2011arXiv1111.4258G} showed that the peak energy is mainly determined by the bulk Lorentz factor of the flow.

Both accreting black holes and rapidly rotating magnetars have been invoked for launching the jet. In the proto-magnetar model for LGRBs \citep[see e.g.][and references therein]{2011MNRAS.413.2031M} the highest magnetised flows give rise to the brightest flows \citep[see Eq. 11 in][and the following discussion]{2011MNRAS.413.2031M}. The energy budget is limited by the magnetar rotational energy \citep{1992Natur.357..472U}, which implies also that less energy remains to power the following X-ray emission. This qualitatively accounts for the anticorrelation between $\epsilon$ and $E_{pk}$. SGRB outflows might be highly magnetised as in the previous scenario, and a limited energy budget is expected also in the merging of two compact objects if these are neutron stars. Although the scaling between $E_{\gamma,iso}$ and $E_{pk}$ is different in this case, the limited energy budget implies again that the more energetic is the prompt emission (the higher is $E_{pk}$), the lower is the energy budget left over after the main event.

The correlation does not include the energy emitted during the flaring activity, which is present in $40\%$ of the GRBs of our sample. The energy content of flares is usually $\sim 3\% E_{\gamma,iso}$, but it can be as high as $E_{\gamma,iso}$ for GRB060526\footnote{The most extreme case is GRB050502B \citep{2005Sci...309.1833B}, with a flare whose fluence is $22.7$ times larger than the fluence of the underlying continuum, and $2.6$ times the fluence observed by BAT.}, and is $\sim 25\%$ of their underlying continuum energy $E_{X,iso}$ (for details see M12). The inclusion of the flare energy into $E_{X,iso}$ does not improve significantly the correlation scatter. This is not surprising since SGRBs, which are the most scattered population in the three-parameter correlation, have no bright flares \citep{2011MNRAS.417.2144M}. 

After a systematic analysis of all the flare properties \citep{chinca10,giantflares10,2011MNRAS.410.1064M,2011A&A...526A..27B}, their origin remains elusive. Their observational properties point to a direct link with the prompt emission (as e.g. in the form of late-time activity of the central engine; \citealt{2006MNRAS.370L..61P,2006ApJ...636L..29P}). However, they may also be produced by late-time dissipation within the outflow \citep[see e.g.][]{2006A&A...455L...5G}. It is therefore unclear if computing them with either $E_{\gamma,iso}$ or $E_{X,iso}$. Either way we note that even the extreme GRB060526, for which $E_X^{FL}\sim E_{\gamma,iso}$, is still consistent with the correlation at $2-\sigma$ level.

\section{Summary and conclusions}

The analysis of M12 revealed a strong connection between the prompt gamma-ray and the late X-ray emission. In particular, we found a universal scaling among $E_{X,iso}$, $E_{\gamma,iso}$, and $E_{pk}$. In the present work we showed that:
\begin{itemize}
\item the three-parameter correlation is shared by SGRBs and LGRBs, regardless on their different progenitors, environments, etc. It is therefore natural to associate this relation to some properties which are shared by the GRB class as a whole (e.g. the properties of the outflow). It includes also low-energetic GRBs. In particular, GRB980425, GRB031203, and GRB061021, which are outliers of the $E_{\gamma,iso}-E_{pk}$ relation, are consistent with the correlation;
\item it is robust, and does not depend on our definition of $E_{X,iso}$;
\item we speculate that the physical origin of such a relation is connected with the outflow Lorentz factor, which might regulate the efficiency of conversion from gamma-rays to X-rays.
\end{itemize}

\section*{Acknowledgments}
We thank Lorenzo Amati for sharing his data before publication. This work is supported by ASI grant SWIFT I/004/11/0.

\label{lastpage}

\end{document}